# The uncertainty measure for q-exponential distribution function

Congjie Ou[1,2], Aziz El Kaabouchi[2], Qiuping A. Wang[2], Jincan Chen[2,3]

[1]College of Information Science and Engineering, Huaqiao University, Quanzhou 362021, China

[2]Institut Supérieur des Matériaux et Mécaniques Avancés du Mans 44, Avenue F.A. Bartholdi, 72000 Le Mans, France

[3]Department of Physics, Xiamen University, Xiamen 361005, China

**Abstract**: Based on the q-exponential distribution which has been observed in more and more physical systems, the varentropy method is used to derive the uncertainty measure of such an abnormal distribution function. The uncertainty measure obtained here can be considered as a new entropic form for the abnormal physical systems with complex interaction. The entropy obtained here also presents non-additive property for two independent subsystems and it will tend to the Boltzmann-Gibbs entropy when the nonextensive parameter $q$->1. It is very important to find that for different systems with any $q$, this entropic form is always concave and the systemic entropy is maximizable.

**Introduction**

In classical statistical mechanics the Boltzmann factor, i.e., $exp(-x)$, is very important to discuss the thermo dynamical properties of the physical systems. This factor is obtained under some ideal conditions, such as there are no other interactions except the elastic collisions between the particles of the system, the number of particle tends to infinity, etc. However, with the development of modern physics, these ideal conditions become more and more imprecise and the classical theory should be generalized. In fact, if the internal interactions are taken into account, some new distributions such as $f \sim [1-(1-\tilde{q})x]^{1/(1-\tilde{q})}$ have been observed in more and more physical systems[1]-[5], where $\tilde{q}$ can be considered as a nonextensive parameter and $x$ is the random variable of the system. This distribution function is called the q-exponential function and the corresponding generalization of classical theory is named nonextensive statistical mechanics since it started by Tsallis[6] in 1988. Obviously the q-exponential function will reduce to the Boltzmann factor if $\tilde{q}$ tends to 1.

On the other hand, the entropy is a very important quantity for the physical systems. According to the maximum entropy principle (MaxEnt), the probability of each microstate of the system can be determined by the entropic form under the constraints of probability normalization condition and the expectations of the observable quantities. The generalization of the Boltzmann-Gibbs entropy ($S_{BG} = -k\sum_i p_i \ln p_i$) is a kernel of the framework of nonextensive statistical mechanics. The most famous nonextensive entropy was proposed by Tsallis and it can

be written as [6]

$$S_{\tilde{q}} = k \frac{1 - \sum_{i=1}^{W} p_i^{\tilde{q}}}{\tilde{q} - 1},\qquad(1)$$

where $k$ is the Boltzmann constant and $p_i$ is the probability of the $i$-th microstate of the whole states $W$ of the system. Eq. (1) will reduce to the Boltzmann-Gibbs entropy at the $\tilde{q} \to 1$ limit. However, the maximization of the Tsallis entropy at the constraints of probability normalization and escort expectation [7] yields

$$p_i = \frac{1}{Z}\left[1 - (1-\tilde{q})\frac{x_i - \bar{x}}{\sum_j p_j^{\tilde{q}}}\right]^{\frac{1}{1-\tilde{q}}},\qquad(2)$$

where $\bar{x} = \sum_i p_i^{\tilde{q}} x_i / \sum_j p_j^{\tilde{q}}$ represents the escort expectation and $Z$ is the partition function. Eq. (2) is a self-referential equation so it is different with the q-exponential distribution function. Hence, it is not sufficient to draw a conclusion that the q-exponential distribution function is a result of maximizing the Tsallis entropy under proper constraints. The maximization of the Incomplete entropy [8] under proper constraints also yields a self-referential distribution function. In reference [7] the authors has got exactly the q-expectation function by maximizing the Tsallis entropy with the unnormalized energy constraint under the standard Lagrange Multiplier method, but the unnormalized expectation ($\bar{x} = \sum_i p_i^{\tilde{q}} x_i$) is in conflict with the expectation of a constant $C_1$, i.e., $C_1 \neq \sum_i p_i^{\tilde{q}} C_1$. Thus, this method is seldom adopted at present. To sum up, all the nonextensive entropies mentioned above can not yield the q-exponential distribution function under reasonable constraints by using the standard Lagrange Multiplier method during the last two decades.

Very recently, a variational relationship (VarEntropy method) between the random variables and the uncertainty measure of the system has been proposed [9] to discuss the probability distribution function of the system. It is worth to point out that the uncertainty measure can be considered as the entropy of the system [10][11] since the "VarEntropy" is exactly the reverse process of "MaxEnt" [12]. In principle, one can get different entropic forms for any distribution function by using VarEntropy method. In the present paper, the entropic form of q-exponential distribution is investigated by using VarEntropy method and some novel properties of this new entropic form are obtained.

**VarEntropy method for q-exponential distribution function**

The variational relationship between a random variable $x$ of the system and the uncertainty measure can be written as [9]

$$dI = d\overline{x} - \overline{dx} = \sum_{i=1}^{W} x_i dp_i, \qquad (3)$$

where " $^-$ " means the expectation or average value of a quantity. The q-exponential distribution can be written as

$$p_i = \frac{1}{Z}[1-(1-\widetilde{q})x_i]^{1/(1-\widetilde{q})}. \qquad (4)$$

Eq. (4) has another symmetrical form

$$p_i = \frac{1}{Z}[1-(q-1)x_i]^{1/(q-1)} \qquad (5)$$

if one set $q = 2 - \widetilde{q}$. The partition function $Z$ can be written as

$$Z = \sum_{i=1}^{W}[1-(q-1)x_i]^{1/(q-1)} . \tag{6}$$

The value of $Z$ is depended on the random variable distribution $\{x_i\}$ and the nonexetensive parameter $q$. If we make an energy shift: $x_i \to x_i - a$, it is possible to get (see the detail in Appendix)

$$Z = \sum_{i=1}^{W}[1-(q-1)(x_i-a)]^{1/(q-1)} = 1, \tag{7}$$

where $a$ is a real number. Substituting Eq. (7) into Eq. (5) yields

$$p_i = [1-(q-1)(x_i-a)]^{1/(q-1)}, \tag{8}$$

and then

$$x_i = \frac{1-p_i^{q-1}}{q-1} + a . \tag{9}$$

From Eqs. (3) and (9) one can obtain the uncertainty measure for the q-exponential distribution function as

$$I = \int dI = \frac{C - \sum_{i=1}^{W} p_i^q}{q(q-1)}, \tag{10}$$

where $C$ is the integral constant. It is worth to point out that for a determined system, if the condition $p_i = \delta_{1,i}$ is chosen, the uncertainty measure should be equal to zero, so one can easily get $C = 1$. Finally, we have

$$I = \frac{1 - \sum_{i=1}^{W} p_i^q}{q(q-1)} . \tag{11}$$

Eq. (11) will tend to $I_{BG} = -\sum_{i=1}^{W} p_i \ln p_i$ at the $q \to 1$ limit. Comparing $I_{BG}$ with the form of the Boltzmann-Gibbs entropy, it is clear that $S_{BG} = kI_{BG}$. For the sake of convenience, we set the Boltzmann constant $k = 1$ below. Eq. (11) can be considered as a new entropic form since it can be maximized under certain constraints to yield the q-exponential distribution function by standard Lagrange multiplier method.

**Discussion**

The certain constraints are both the probability normalization and the energy expectation and can be written as $\sum_{i=1}^{W} p_i = 1$ and $\sum_{i=1}^{W} p_i \varepsilon_i = U$, where $U$ is the energy of the system. Thus, the Lagrangian can be written as

$$L = I - \alpha \sum_i p_i - \beta \sum_i p_i \varepsilon_i, \qquad (12)$$

where $\alpha$ and $\beta$ are two Lagrange multipliers. The maximization of entropy means the derivative of Eq. (12) with respect to $p_i$ should be equal to 0, i.e.,

$$\delta L \big|_{p_i} = 0, \qquad (13)$$

and then

$$\begin{aligned} p_i &= [(1-q)\alpha - (q-1)\beta \varepsilon_i]^{\frac{1}{q-1}} \\ &= [1 - (q-1)(x_i - a)]^{\frac{1}{q-1}} \end{aligned}, \qquad (14)$$

where $x_i = \beta \varepsilon_i$ and $a = \dfrac{1}{q-1} - \alpha$. Eq. (14) is nothing but the q-exponential function so it can be clearly seen that the MaxEnt principle and the VarEntropy method are exactly self-consistent if we take the entropic form of Eq. (11) and the distribution function of Eq. (5) into account. It is important to note that Eq. (11) is the first entropic form that will yield exactly the non-self-referential q-exponential distribution which is coincide with many experimental observations in abnormal physical systems.

Let us assume that a composed system C consists of two independent subsystems A and B, which have the probability distribution $\{p_i(A)|i=1,2,...W_A\}$ and $\{p_j(B)|j=1,2,...W_B\}$, respectively. The independence between the subsystems means that $p_{i,j}(A|B) = p_i(A)p_j(B)$ and then the entropy of the total system C can be written as

$$I(C) = I(A) + I(B) - q(q-1)I(A)I(B). \qquad (15)$$

The third item in the right hand side of Eq. (15) means the nonextensivity of this new entropic form. It will disappear if the nonextensive parameter $q \to 1$.

The another property of Eq. (11) is the concavity. Let us consider a system which an entropic form described by Eq. (11) and only has two microstates, the probability of state 1 is $p_1$ and the other is $(1-p_1)$. As it is shown in Fig. 1, all the curves are concave. It means that this entropic form can be maximized for different physical systems with arbitrary $q$. In Fig. 1 we have not presented the cases of $q>1$ since there doesn't exist a real number solution for $a$ in Eq. (7) sometimes. This constraint condition is illustrated by Eq. (A12) in Appendix. Nevertheless, for a system with given energy spectrum and total number of microstates, if there exist some values of $q$ ($q>1$) that can satisfy Eq. (A12) one can also obtain the entropy curves with respect of the probability variation. The entropies of different $q$ values also keep the concavity, as shown in Fig. 2.

**Conclusions**

By using the VarEntropy method we obtain a new entropic form from the q-exponential distribution function which has been observed in more and more complex physical systems presenting long range interactions and/or long-duration memory. This entropy can yield *exactly* the q-exponential distribution by standard Lagrange multiplier method under the probability normalization and expectation constraints. It can be considered as a nonextensive generalization of the traditional Boltzmann-Gibbs entropy since the traditional one can be covered at the $q \to 1$ limit and the nonextensive item in Eq. (15) disappears in this case. It is very significant to point out that the entropy is always concave with an arbitrary nonextensive parameter *q*. It means that it is reasonable to adopt this entropy form to evaluate the uncertainty (or information) of a system since it is maximizable no matter what *q* the system has.

## Acknowledgements

This work has been supported by the National Natural Science Foundation (No. 10947114), People's Republic of China, and by the Fujian Natural Science Foundation (No. 2010J05007), People's Republic of China, and by the Science Research Fund (No. 07BS105), Huaqiao University, People's Republic of China.

the variational entropy form ", Phys. Lett. A (2010) in press, arXiv: 1007.4729

[13]   Wang Q A   2001 Chaos, Solitons & Fractals **12** 1431

Appendix

In order to prove that a real number $a$ may be found to satisfy Eq. (7), we introduce a new function as

$$f(a) = \sum_{i=1}^{W} [1-(q-1)(x_i - a)]^{1/(q-1)} . \tag{A1}$$

In order to guarantee $f(a) \in R$,

$$1-(q-1)(x_i - a) \geq 0 \tag{A2}$$

must be satisfied. Below, we will discuss the two cases of $0 < q < 1$ and $1 < q$, respectively.

1) The case of $0 < q < 1$

In this case Eq. (A2) can be written as

$$a \leq x_i - \frac{1}{(q-1)} . \tag{A3}$$

Since $x_i$ is one of the random values among $W$ microstates of the system, one can get the upper limit of the parameter $a$ as

$$a_{max} \leq x_{min} - \frac{1}{(q-1)}, \tag{A4}$$

where $x_{min} = \min\{x_i\}$. On the other hand, the derivative of Eq. (A1) with respect to $a$ gives

$$f'(a) = \sum_{i=1}^{W} [1-(q-1)(x_i - a)]^{\frac{1}{q-1}-1} \tag{A5}$$

It is easily seen from Eqs. (A2) and (A5) that $f'(a)$ is always positive. It means that $f(a)$ is a monotonically increasing function of $a$ in the region $(-\infty, x_{min} - 1/(q-1))$, as shown in Fig. A1, where

$$\lim_{a \to -\infty} f(a) = 0 \tag{A6}$$

and

$$\lim_{a \to x_{\min}-1/(q-1)} [1-(q-1)(x_{\min}-a)]^{\frac{1}{q-1}} \to +\infty, \quad \lim_{a \to x_{\min}-1/(q-1)} f(a) \to \infty \tag{A7}$$

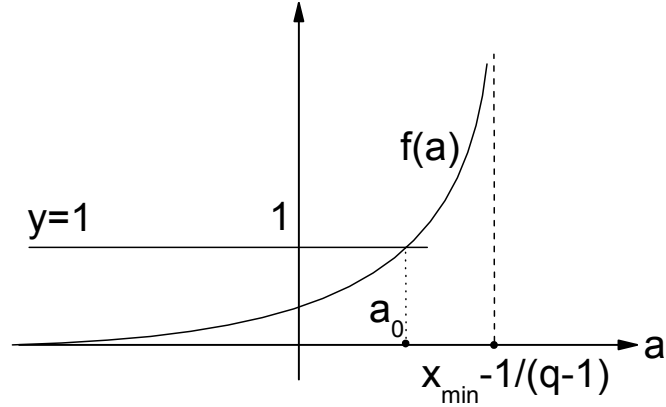

Fig. A1 The schematic diagram of $f(a)$ versus a for the case of $q<1$

It is clearly seen from Fig.A1 that the curve of $f(a)$ has a crossing point with the straight line $y=1$. Thus, one can draw a conclusion that Eq. (8) has a real number solution $a_0$ marked in Fig.A1.

2) The case of $1<q$

When $1<q$, Eq. (A2) can be written as

$$a \geq x_i - \frac{1}{(q-1)} \tag{A8}$$

In such a case, the lower limit of $a$ is determined by

$$a_{\min} \geq x_{\max} - \frac{1}{(q-1)} \tag{A9}$$

where $x_{\max} = \max\{x_i\}$. It can be easily seen from Eqs. (A2) and (A5) that when $1<q$, $f(a)$ is also a monotonically increasing function of a in the region $(x_{\max}-1/(q-1),+\infty)$, as shown in Fig. A2, where

$$\lim_{a \to +\infty} f(a) \to +\infty \tag{A10}$$

and

$$\lim_{a \to x_{max}-1/(q-1)} f(a) = \sum_{i=1}^{W}[(q-1)(x_{max}-x_i)]^{\frac{1}{q-1}}$$
$$\leq \sum_{i=1}^{W}[(q-1)(x_{max}-x_{min})]^{\frac{1}{q-1}} = W[(q-1)(x_{max}-x_{min})]^{\frac{1}{q-1}} \tag{A11}$$

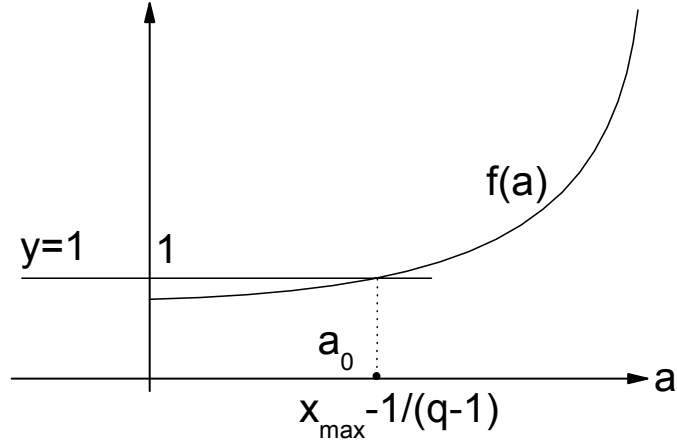

Fig.A2 The schematic diagram of $f(a)$ versus $a$ for the case of $1 < q$

It is seen from Fig. A2 that the curve of $f(a)$ has a crossing point with the straight line $y = 1$ under the condition

$$W[(q-1)(x_{max}-x_{min})]^{\frac{1}{q-1}} \leq 1 \tag{A12}$$

This indicates that when Eq. (A12) is satisfied, we can have a real number solution $a_0$ for Eq. (7).

Fig captions:

Fig 1. The entropy of two states system varies with the probability for the cases of $q \leq 1$.

Fig 2. The entropy of two states system varies with the probability for the cases of $q > 1$.

Fig 1.

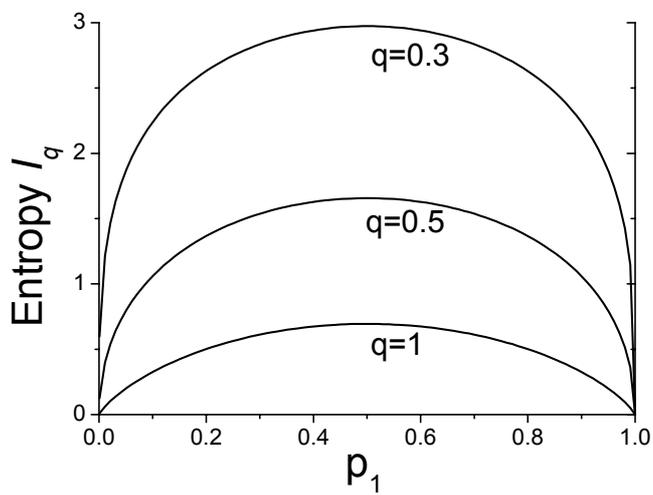

Fig. 2

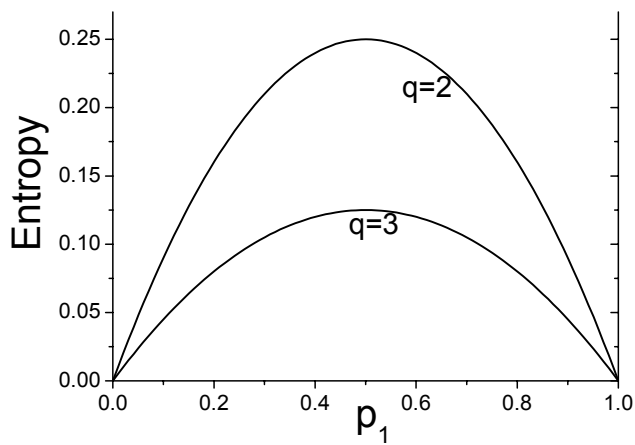